# Location and Position Estimation in Wireless Sensor Networks

Muhammad Farooq-i-Azam
*COMSATS Institute of Information Technology, Lahore, Pakistan*

Muhammad Naeem Ayyaz
*University of Engineering and Technology, Lahore, Pakistan*

**ABSTRACT**

*A wireless sensor network comprises of small sensor nodes each of which consists of a processing device, small amount of memory, battery and radio transceiver for communication. The sensor nodes are autonomous and spatially distributed in an area of investigation. Certain applications and protocols of wireless sensor networks require that the sensor nodes should be aware of their position relative to the sensor network. For it to be significant and to be of value, the data such as temperature, humidity and pressure, gathered by sensor nodes must be ascribed to the relative position from where it was collected. For this to happen, the sensor nodes must be aware of their relative positions. Traditional location finding solutions, such as Global Positioning System, are not feasible for wireless sensor nodes due to multiple reasons. Therefore, new methods, techniques and algorithms need to be developed to solve the problem of location and position estimation of wireless sensor nodes. A number of algorithms and techniques based upon different characteristics and properties of sensor nodes have already been proposed for this purpose. This chapter discusses the basic principles and techniques used in the localization algorithms, categories of these algorithms and also takes a more closer look at a few of the representative localization schemes.*

## 1. INTRODUCTION

A Wireless Sensor Network is a network of tiny sensor nodes which communicate with each other through a wireless communication link. Each sensor node typically consists of a processing device, small amount of memory, battery and radio transceiver for communication. These sensor nodes obtain data, e.g. temperature, pressure and humidity, do some local processing, and transmit the data to a neighbor node or a beacon node, which, in turn, may be connected to a central computer where major processing is performed. As is evident, this central computer may be part of a bigger computer network so that the information can be communicated from this central computer to other computers which are part of the bigger network.

The wireless sensor networks can be used in diverse applications in both industrial and commercial environments. Some of the most commons applications of wireless sensor networks include object tracking, habitat monitoring, fire detection, traffic monitoring and area monitoring. Some of the typical characteristics of sensor networks are small sized nodes, mobile nodes, a dynamic network topology, harsh operating environments and limited energy or power resource that these nodes should utilize efficiently as these may remain in an area for years without more energy being available.

There are many challenges involved at various levels and stages in the development of wireless sensor networks as discussed by Akyildiz et al. in [1]. For example, the physical layer design of a wireless sensor node, which should be very small and yet accommodate all the functions that it is required to perform, presents many challenges. Similarly, new algorithms and protocols from link layer to application layer need to be developed. New operating systems to run on such tiny nodes are needed and to write these tiny operating systems, there is a need to develop new programming languages and new programming paradigms. One of the important problems is the localization of sensor nodes i.e. determination of positions of nodes in the sensor field. This is important due to various reasons. For example, the data collected by a sensor node must be ascribed to the location from where it was collected. The data would not be useful if the location where it belongs to is not known. The set of values of temperature and humidity, for instance, collected by the sensor nodes are not meaningful unless the respective position coordinates from where these values were recorded are known.

An example of wireless sensor network application where location information is important is target tracking. Likewise, in a sensor network meant for earthquake disaster relief, the sensor positions must be known to ascertain the location of survivors buried somewhere in the rubble of a collapsed building. Similarly, one of the biggest challenges in sensor networks is the efficient utilization of energy resource which is not easily available to sensor nodes. And one of the most energy dependent operations is data transmission from sensor nodes to base stations which should use some energy-efficient and energy-aware routing algorithm. One of the approaches being worked out and which holds great promise is geographic location based routing, which is based upon mathematical modeling of sensor positions instead of using IDs. Again for location-based approach to be possible, the locations of the nodes must be known.

Due to various constraints, existing localization systems, such as GPS, cannot be used for the localization of wireless sensor nodes. Therefore, new strategies and algorithms for the localization of sensor nodes are needed to be designed and developed. The algorithms should be designed within the constraints defined, the characteristics of the sensor nodes and sensor network. In wireless sensor networks using only static sensor nodes, localization algorithm usually runs only at the time of initial deployment of the nodes. However, in a sensor network using mobile sensor nodes, the localization algorithm is needed to determine the new positions of mobile nodes as they move in the sensor field. Hence, localization algorithms for mobile sensor nodes need more energy compared to algorithms designed for static sensor nodes.

## 2. BACKGROUND

Certain applications of wireless sensor networks require that the sensor nodes should be aware of their position relative to the sensor network. For it to be significant and to be of value, the data such as temperature, humidity and pressure gathered by sensor nodes must be ascribed to the relative position from where it was collected. For this to happen, the sensor nodes must be aware of their positions. The literature has come to term this problem of location or position estimation of sensor nodes simply as *localization*. The term localization has earlier been used in robotics where it is used to refer to determination of location of a mobile robot in some coordinate system. Under certain circumstances, the nodes should not only by aware of their position but also the direction or orientation relative to the network [8].

In a sensor network, the nodes may be categorized as:

*Dumb Node (D)*

It is the node that does not know its position and which would eventually find its location and position from the output of the localization algorithm under investigation. Dumb nodes are also known as *free* or *unknown* nodes.

*Settled Node (S)*

A settled node is a node which was initially a dumb node but managed to find its position using the localization algorithm.

*Beacon Node (B)*

A beacon node is a node that knows its position from the very start and always knows its position afterwards also without the use of localization algorithm. It has a mechanism other than the localization algorithm to find its position. For example, the beacon node may be equipped with a GPS device or it may be placed at a position with known coordinates. The beacon nodes are also called *reference nodes*, *anchor nodes* or *landmark nodes*.

It should be noted that sensor nodes may have *symmetric* or *asymmetric* communication links. If two nodes *u* and *v* are symmetric then *u* reaches *v* and *v* reaches *u* as well. In the case of asymmetric communication links, either *u* reaches *v* or *v* reaches *u* but both *u* and *v* do not reach each other simultaneously.

Let us now consider a sensor network which is symmetric, two-dimensional and arranged in a square shape. Then this sensor network can be represented as a graph *G(V, E)* where the set of sensor nodes can be represented as set of vertices as under:

$$V = \{ v_1, v_2, ..., v_n \}$$

The set of edges *E* in the graph *G(V, E)* comprises of all edges $e = (i, j) \in E$ iff $v_i$ reaches $v_j$ i.e. the distance between $v_i$ and $v_j$ is less than *r* where *r* is the maximum distance between the two nodes after which communication between them ceases to exist i.e. if the distance between two nodes is greater than *r*, no direct communication between them is possible. In other words, if the distance between two nodes is greater than *r*, the two nodes are not *neighbor* nodes. The distance between two neighbor nodes $v_i$ and $v_j$ is defined as the weight $w(e) \leq r$ of the edge $e = (i, j)$ between them.

It is to be noted that problem of localization is usually solved only for two dimensions with the supposition that when needed or deployed, it could be extended to three dimensions. It is for this reason, we have stated graph *G(V, E)* to be two-dimensional. Therefore, it can be stated that *G* is a Euclidean graph in which every sensor node has a coordinate $(x_i, y_i) \in \Re^2$ in a two-dimensional space. The coordinate $(x_i, y_i)$ represents the location of a node *i* in the given sensor field.

The sensor node localization problem can now be stated as following:

*Let there be a multihop sensor network represented by a graph G = (V, E). The graph has a set of beacon nodes B with known positions given by $(x_b, y_b)$ for all $b \in B$. The localization problem requires to find the position set $(x_d, y_d)$ of as many dumb nodes $d \in D$ as possible. Finding the location of a node implies finding its latitude, longitude and altitude.*

Problem of node localization and positioning in a sensor network can be solved if each node is equipped with a GPS device. However, in the case of sensor networks, this is not a feasible option for a number of reasons such as:

- GPS receiver and the protocols used are not designed to be energy efficient or energy aware. In the case of sensor networks, energy is a scarce resource and the sensor nodes may be deployed without any sort of battery being replaced for many years. Therefore, GPS devices are not suitable for the solution of localization problem in wireless sensor networks. It is, nevertheless, possible that *beacon nodes which constitute only a fraction of the total number of nodes are equipped with a GPS device* so that these can serve as reference nodes for other nodes to solve the problem of their position awareness using the localization algorithm.

- GPS devices are much more expensive. If these are added to every sensor node in the network somehow, cost of deployment may increase to an extent so as to render the sensor network solution unfeasible for a particular problem.

- One of the required properties of sensor nodes is that these should be of very small size. With the addition of GPS device, the size of sensor nodes would become quite large which again violates one of the primary requirements of a sensor node.

- GPS devices depend upon satellites for their functioning. In cases or under circumstances when no satellite link is available, GPS ceases to function. In certain applications, this really can be the case e.g. indoor applications and Mars exploration.

Due to the above reasons, GPS devices are normally used only in a fraction of nodes which serve as reference nodes to solve localization problem of other nodes. Such nodes are also called beacon nodes. Alternatively, it is possible to avoid use of GPS altogether by positioning a few nodes at fixed points so that their position is known a-priori so that these nodes can serve as beacon nodes. The sensor field can then exploit inherent radio frequency (RF) capabilities of the sensor nodes or some other techniques to determine their positions using a localization algorithm. The accuracy with which the dumb nodes can determine their location depends upon the transmission range of beacon nodes and distance between two adjacent beacon nodes i.e. density of beacon nodes.

Various algorithms for the localization of wireless sensor networks have been proposed and this is currently a hot area of research with many recent publications such as by Amundson & Koutsoukos [4], Mao & Fidan [15], Pal [17], Wang & Xu [31], Zhang, Foh, Seet & Fory [35] and Kim & Kim [14].

## 2.1 Classification of Localization Algorithms

Majority of the existing localization algorithms may be classified as *ranged-based* or *range-free* depending upon whether the algorithm uses distance estimation or some other information for estimating the node locations. Range-based algorithms usually use sensor field geometry information to determine node locations. Communication between beacon nodes and dumb nodes also helps determine this geometric information about their relative placement e.g. distance between the two nodes or the angles of a triangle formed by the beacon nodes. This information is then further used to determine node location. When distance is used as a primary means to determine node location, this is termed as *lateration*, and when angle information is used for localization, it is known as *angulation*. For node localization in a plane, precise distance measurements from at least three beacon nodes are required and we use *trilateration* for position estimation of a node. Intersection of three circles around the three beacon nodes gives a single point as position of the node as is shown in Figure 1.

The same technique can be extended to three-dimensional space by the addition of a fourth beacon node. However, in actual practice, distance measurements are seldom precise and intersection of three circles may result in more than one point. The scheme may be improved by employing more than three beacon nodes for a plane and we then use *multilateration* to calculate the node position.

In the *angulation* technique, angle information is used to deduce position of the node. Two beacon nodes and a dumb node form the vertices of a triangle and the lines joining them form the sides of the triangle. As the positions of beacon nodes are known, the distance between them, that is, one side of the triangle is known. If the two angles that the dumb node forms with the two beacon nodes are measured, the location of the dumb node can be calculated as third point of the triangle. This method of determining location of a node is termed as *triangulation*.

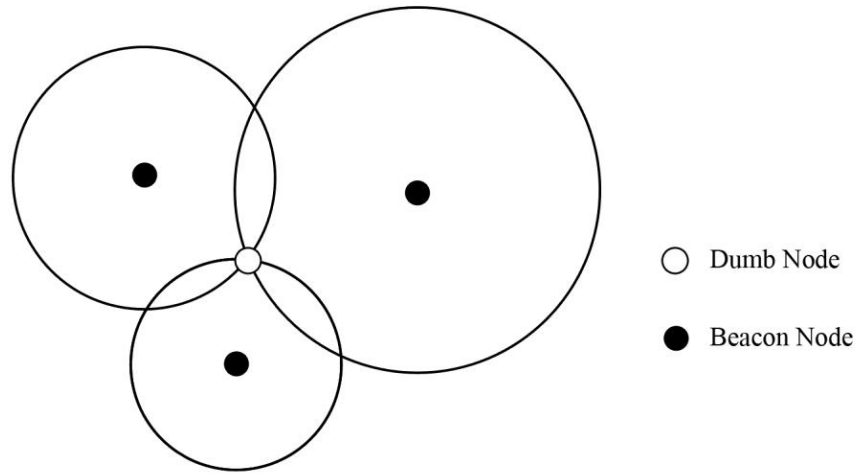

Figure 1. Trilateration – Intersection of three circles around three beacon nodes gives position of the dumb node.

In *range-free* localization algorithms, a node determines its position merely by finding the beacon nodes in its proximity and for this reason, range-free algorithms are also termed as *proximity-based* or *connectivity-based* algorithms. Such algorithms usually provide coarse-grained localization. However, with sufficient number of beacon nodes with overlapping transmission regions, a more accurate localization is possible. Range-free algorithms are robust against the fluctuations of the wireless channel as the decision whether a node is in proximity of another node is based upon connectivity information sampled over a long period of time. Hence, short and temporary variations in the wireless channel do not affect the accuracy of location estimation. A range-free algorithm calculates the position without having to find the distance between the sensor nodes. A range-based localization algorithm is more accurate but has major computational cost and usually additional hardware and hence increased energy requirements. On the other hand, a range-free algorithm is less accurate but does not require additional hardware and has smaller computational overhead.

Localization algorithms may also be classified either as *centralized* or *distributed* depending upon whether they use central processor to estimate the positions of all nodes or else nodes performs local processing to determine their positions. In the case of centralized approach, node positions are calculated at a central processor. To calculate positions of distant nodes, the central unit usually needs certain parameters from these nodes, which depend on the particular localization algorithm. These parameters from the distant nodes are sent to the central unit using the transceiver unit. The central unit computes positions of all nodes and sends the results back to them. As is evident, this approach may involve a lot of communication overhead for the dumb nodes, which, in turn, would require these nodes to have more battery power for sending and receiving the parameters. The centralized approach also introduces a single point of failure. If the central unit fails for some reason, the entire localization process failsk. This approach is also prone to additional delays due to propagation delays involved in the transmission of parameters and results and the large amount of processing involved at the central unit. Furthermore, central processing approach is not suited to the very nature of ad-hoc networks.

In decentralized or distributed approach, the dumb nodes determine their position themselves by performing local processing. The dumb nodes usually need information from neighbor nodes and beacon nodes to be able to determine their position. For decentralized approach to work, the dumb nodes must possess local processing capability. However, communication overhead is less compared to the centralized approach.

Localization algorithms may also be classified as *fine-grained* or *coarse-grained*. Some applications of sensor networks need to determine only symbolic location of a sensor node e.g. whether sensor node is inside room A or room B. This type of symbolic estimation of node position is termed as coarse-grained localization. In some other applications, we need to determine a more accurate estimate of node position e.g. node X is located at coordinates (x, y). This type of node localization is termed as fine-grained localization.

Another classification of the localization algorithms depends upon whether the algorithm is designed for *outdoor* unconstrained environment or for the *indoor* constrained environment. An earliest localization algorithm [6] for the outdoor environment is due to Bulusu, Heidmann & Estrin.

Yet another way to classify localization algorithms is to consider whether the algorithm is designed for a sensor field in which the sensor nodes are *fixed* or for a sensor field comprising of *mobile* sensor nodes. Majority applications of wireless sensor networks use static nodes. As a result, many of the localization algorithms consider sensor networks comprising entirely of static nodes only. However, a few applications of wireless sensor networks deploy mobile nodes and hence a few localization algorithms, such as the one by Kim & Kim [14], take the mobile nature of the nodes into consideration.

## 2.2 Characteristics of Localization Algorithm

The main objective of a localization algorithm is to determine position of a node. However, there are certain criteria that the algorithm should meet for it to be practicable. The criteria usually depend upon the type of application for which the localization algorithm is designed. General design objectives or desired characteristics of an ideal localization algorithm are:

- It is highly desirable that the localization algorithms are *RF-based*. The sensor nodes are equipped with a short-range RF transmitter. An efficient localization algorithm exploits this radio capability for localization in addition to its primary role of data communication.

- A wireless sensor network is ad hoc in nature. The localization algorithm should take the *ad hoc* nature of the network into consideration.

- The nodes should be able to determine their position in as small time as possible so that the localization algorithm has a *low response time*. This would enable sensor nodes to be deployed quickly.

- The position of the sensor node found by such an algorithm should be *accurate* enough for the specific application for which this algorithm is being used.

- The algorithm must be *robust* so that it may work in adverse conditions.

- The algorithm should be *scalable* so that if sensor nodes are added or removed, it should still be able to work out the position of the nodes. Furthermore, the algorithm should produce acceptable results for sensor networks comprising of small to large number of nodes.

- The localization algorithm should be *energy efficient* and preferably *energy aware* as well because the sensor nodes are autonomous and normally do not have any external source of power.

- The localization algorithm should be *adaptive* to the change in the number of beacon nodes. If the number of available beacon nodes changes, the algorithm should still be able to provide location estimates. However, the accuracy of node estimates will change with the change in number of available beacon nodes. In general, with a higher number of beacon nodes, a localization algorithm is able to compute more accurate estimates of node positions.

- The algorithm should be *efficient* so that it is able to compute node locations with as small number of beacon nodes as possible.

- The algorithm should be *universal* so that it is able to compute node locations under all conditions of changing environments and weather. In particular, it should work in constrained environments such as indoors and unconstrained environments such as outdoors.

Only an *ideal* localization algorithm will be able to meet all the goals stated above. The localization algorithms in practice will only meet a subset of these characteristics depending upon the particular application for which it is designed.

## 3. DISTANCE ESTIMATION

Distance estimation between two nodes is an important function performed by range-based algorithms. A range-based algorithm estimates the position of a sensor node by using the distance information between the nodes which, in turn, is calculated using some physical measured quantity.

The distances between dumb nodes and the beacon nodes are usually determined by adding some additional hardware to the nodes or by using the existing radio communication facility on the sensor nodes. Certain characteristics of wireless communication between dumb and beacon nodes are determined by the distance between them. If these characteristics are quantified and measured at the receiving sensor node, these can be used to estimate the distance between the nodes. The characteristics generally used for this purpose are:
- Received Signal Strength Indicator (RSSI)
- Time of Arrival (ToA)
- Time Difference of Arrival (TDoA)
- Angle of Arrival (AoA)

### 3.1 Received Signal Strength Indicator (RSSI)

A signal is attenuated as it travels from transmitter to receiver. Longer the distance the signal has to travel, greater is the attenuation. Therefore, strength of the received signal can be used to estimate the distance between the transmitter and receiver. The distance can be calculated using the following information:

- Transmitted power of the signal
- Received power of the signal i.e. received signal strength
- Path loss model

Using these three parameters, power of the received signal $P^{ij}_R$ transmitted by node *i* and received at node *j* at time *t* can be expressed as:

$$P^{ij}_R(t) = P^{i}_T - 10\eta \log(d_{ij}) + X_{ij}(t)$$

In this equation:

$P^{ij}_R(t)$ is power of the received signal at receiver node *j* transmitted by node *i* at time *t*.

$P^{i}_T$ is transmitted power of signal transmitted by node *i*.

$\eta$ is attenuation constant, value of which depends upon the surroundings of the receiver node *j*.

$d_{ij}$ is distance between transmitter node *i* and receiver node *j*.

$X_{ij}(t)$ is uncertainty factor or channel model whose value depends upon multipath fading and shadowing.

Equation above can be solved for distance $d_{ij}$ between beacon node *i* and receiver node *j* as all other parameters are known. Distance estimation using this technique is quite attractive due to the following factors:

- As the sensor nodes communicate with each other for data transfer, the solution to the problem of localization using RSSI values is only an added benefit. It does not need any additional hardware circuitry and no size and weight is added to sensor nodes which should remain as tiny as possible.

- Localization based upon RSSI also saves power which should, otherwise, be needed by the additional hardware used for localization.

Apart from the merits of an RSSI based localization scheme, there are some demerits as well. Some of these are described below:

- The fundamental assumption in a localization scheme based upon RSSI is that the signal suffers from the same amount of attenuation for the same distance travelled. However, in actual practice, this is not always the case due to factors such as multipath fading, fast fading and shadowing. Savarese, Rabaey & Langendoen [23] have reported that ranging errors of the order of ±50% are possible even when both the transmitter and receiver nodes are stationary. This problem can be remedied, to some extent, by taking more measurements. In addition, statistical techniques may be employed to filter out incorrect values as suggested by Ward, Jones & Hopper [33].

- If there are obstacles between the beacon node and dumb node such that line of sight communication between them is not possible, the later receives signal transmitted by the beacon node after reflection from the surroundings. As a result, the signal suffers far greater attenuation than the case of line of sight communication between the two nodes, and as a consequence, the RSSI value is not a true indicator of the distance vector between the two nodes and yields a distance estimate which is much greater than the actual distance [5]. The resultant error cannot be corrected by increasing the number of measurements since the additional measurements are still based upon RSSI values of signals which are received after reflection from the surroundings.

- As the values of channel model $X_{ij}$ and attenuation constant $\eta$ depend upon the surroundings, an estimation of these parameters is needed using calibration before sensor nodes are deployed in the sensor field. Under certain circumstances, such as deployment in a battlefield, this is not always possible.

- As suggested by Whitehouse & Culler [34], the transceiver units in all sensor nodes should be calibrated so that the RSSI values correspond to the actual strength of the received signal.

## 3.2 Time of Arrival

This technique of distance estimation uses the following relationship that relates the distance travelled by a signal to the time taken provided that the speed of propagation is known.

$$d = v \times t$$

where *d* is distance, *v* is speed of the signal and *t* is time taken by the signal to travel the distance *d*. Therefore, if the time taken by a signal to propagate from the beacon node to the dumb node, which is called *time of arrival* or *time of flight* is measured and speed of propagation of the signal is known, the distance and hence position of the dumb node can be calculated.

There are two variations of the time of arrival technique:

- One-way time of arrival
- Two-way time of arrival

Furthermore, either of the above techniques may use an RF signal or an ultrasonic pulse for distance estimation.

*3.2.1 One-way time of arrival*

In one-way time of arrival technique, the propagation time of one-way trip of the signal from the beacon node *i* to dumb node *j* is measured. This is given by the difference between the sending time $t_i$ at beacon transmitting node and receiving time $t_j$ at the receiving dumb node. The distance $d_{ij}$ between the two nodes *i* and *j* is then given by:

$$d_{ij} = v \times (t_j - t_i)$$

With this approach, the receiver node calculates its position in a secure manner without disclosing its location information to the transmitting node, and hence is also termed as *passive time of arrival* localization. Obviously, transmitting node is usually a beacon node and the receiving node is the dumb node. As stated earlier, an RF signal or ultrasonic pulse can be used for distance estimation using this technique.

It is to be noted that for the one-way time of arrival technique to work the receiver must know the time of transmission of the signal. In the case of an RF signal, the transmitting node can embed this information in the beacon signal that it sends to the receiver. However, RF signals travel at a very high speed, which is almost equal to the speed of light i.e. $3 \times 10^8$ m/s. Their use for distance estimation using time of flight requires extremely accurate and stable clocks, and highly precise hardware for time measurement. For example, if sensor nodes are located 10 meters apart and RF signal is used for ranging, then the time taken by the signal to travel from the transmitting node to the receiving node is given by:

$$t = d / v = 10 / 3 \times 10^8 = 3.33 \times 10^{-8} = 33.3 \ ns$$

If the transmitter and receiver clocks are not tightly synchronized and are out of sync by even 1 ns, then it will result in an error of the order of 1 ns in time of arrival measurement, which will reflect in distance estimation error of $d_e = 1 \times 10^{-9} \times 3 \times 10^8 = 0.3$ meter. The requirement of highly stable and accurate clocks and precise timing measurement means addition of complex and costly hardware to the sensor node and necessitates usage of time synchronization algorithm along with localization algorithm so as to increase size and processing need and thereby energy consumption. As a result, one-way time-of-arrival measurements using RF signals are not considered as a choice for the existing and near future sensor hardware. However, in some media, other than air, e.g. under water and earth, the propagation speed of signals is reduced and hence can be used for distance estimation using time of flight technique. Furthermore, in many applications of wireless sensor networks, such as monitoring applications, not only do we need to know the location *where* an event occurred, but also *when* the event occurred i.e. its timing information. For example, if a sensor node samples temperatures in a sensor field, one might be interested in knowing both the position and timing information of the temperature values. For a sensor network to be able to provide this timing information, the sensor nodes must be accurately synchronized. A localization algorithm can exploit this time synchronization to estimate node position. In fact, some proposed algorithms, such as Synapse by De Oliveira, Nakamura, Loureiro & Boukerche [9], combine time synchronization and localization problems so as to propose a single solution of localization in time and space.

Ultrasonic waves have speed far smaller than RF signals e.g. speed of an ultrasonic wave in the air at 20 ºC is 343.26 m/s. As a result, the hardware required for timing measurement is not complex. However, using ultrasonic pulses for distance estimation has its own drawbacks, some of which are described below:

- Use of ultrasonic pulses for distance estimation will necessitate addition of extra hardware on the sensor nodes for generation and reception of ultrasonic pulses. This will affect the size, cost and energy-efficiency of the sensor node.

- Speed of an ultrasonic wave changes with temperature. As a result, if there are large variations of temperature in the field of sensor network deployment, the distance estimation will change with change in temperature. As a remedial measure to this problem, extra pieces of hardware need to be incorporated on sensor nodes, which again affects the size, cost and energy-efficiency of the sensor nodes.

*3.2.2 Two-way time of arrival*

In the two-way approach, the receiver node sends the signal back to the transmitter node which measures the round-trip time for distance estimation between transmitter and receiver. Suppose sensor node *i* transmits signal at its local time $t_{i1}$. It reaches sensor node *j* at its local time $t_{j1}$. After some delay, the sensor node *j* sends the signal back at its local time $t_{j2}$. The signal is received back at node *i* at its local time $t_{i2}$. This is illustrated in Figure 2.

Now

Total round-trip time including delay = $t_{i2} - t_{i1}$

Delay suffered at $j$ = $t_{j2} - t_{j1}$

Actual roundtrip time = $(t_{i2} - t_{i1}) - (t_{j2} - t_{j1})$

One-way time of flight = $\dfrac{(t_{i2} - t_{i1}) - (t_{j2} - t_{j1})}{2}$

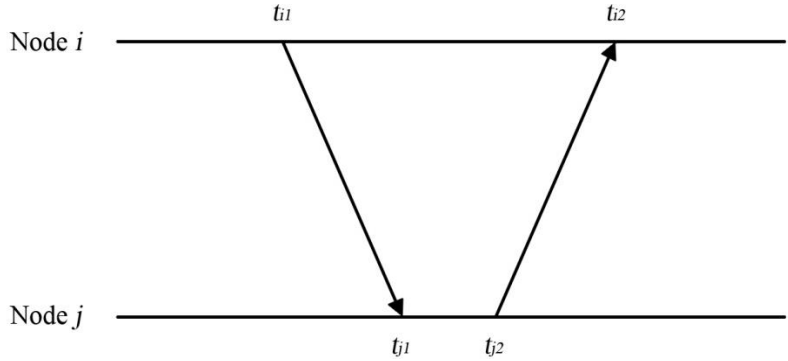

Figure 2. Signal propagation times in two-way time of arrival distance estimation technique.

Hence, the distance $d$ between the two nodes is given by:

$$d = \dfrac{(t_{i2} - t_{i1}) - (t_{j2} - t_{j1})}{2} \times v$$

It should be noted that node $i$ subtracts its local time $t_{i2} - t_{i1}$ to get the total roundtrip time. Similarly, node $j$ subtracts its local time $t_{j2} - t_{j1}$ to get the processing delay. As both nodes have to process only their respective local times, no time synchronization between the nodes is necessary. This saves extra hardware cost and energy required for time synchronization.

With the two-way approach, the beacon node, which is the transmitting node, has to carry out processing for distance and position estimation. The result is then sent back to the dumb node. As there are a large number of dumb nodes per beacon node, a particular beacon node may have to process localization for many dumb nodes. This will increase processing overhead on the beacon nodes by manifold. Furthermore, communication overhead is also involved as the beacon node has to send the result back to the dumb node.

With the one-way time of arrival approach, the beacon nodes do not have any processing overhead as localization processing is carried out by the dumb nodes. In this way, the processing task is uniformly distributed amongst all the dumb nodes, and beacon nodes are also free to send timely beacons to the dumb nodes.

### 3.3 Time Difference of Arrival (TDoA)

Time difference between the receiving of two signals at a node is easier to measure compared to time of arrival of a signal. This time difference information can then be used to estimate the distance between the two nodes. Advantage of using time difference instead of time of arrival is that errors in time difference measurement are tolerable and do not have a pronounced effect on the accuracy of estimation of distance between two nodes. As a result, the hardware required for time measurements is less complex and less costly and hence the method is also efficient in terms of energy consumption.

The TDoA techniques can be classified into two main categories [2]:

- Multi-node TDoA

- Multi-signal TDoA

*3.3.1 Multi-node TDoA*

At least three beacon nodes $B_1$, $B_2$ and $B_3$ transmit signals at exactly the same time. Time differences amongst the arrival of these three signals at the receiving dumb node D are measured. The difference from a pair of beacon nodes, say $B_1$ and $B_2$, defines a branch of hyperbola on which the dumb node D is located. Similarly, difference from the pair of beacon nodes $B_2$ and $B_3$ will again give branch of a second hyperbola. The receiving dumb node should lie on this second hyperbola as well. The point of intersection of the two hyperbolas gives the location of the dumb node D.

It should be noted that the nodes should be time synchronized with stable clocks for them to be able to transmit the beacon signals at exactly the same time. This type of TDoA is quite old and was used in the classical long range navigation systems such as LORAN.

*3.3.2 Multi-signal TDoA*

In the time of arrival (ToA) technique using RF signals, sophisticated hardware is needed for precise measurements of time and the nodes should also be time synchronized. One way to alleviate this problem is the use of an ultrasonic signal along with an RF signal. The beacon transmitting node *i* transmits the two signals simultaneously or after some fixed time interval $t_{i2} - t_{i1}$ as shown in Figure 3.

Due to a large difference in their propagation speeds, the RF signal is received first by the receiving dumb node $j$, which records the arrival time $t_{j1}$ of the RF signal. As the speed of ultrasonic signal is very small compared to the speed of RF signal, this time of arrival of RF signal is treated as the time of transmission of the ultrasonic signal. After receiving RF signal, the receiving dumb node prepares itself to receive the ultrasonic signal. Time of arrival $t_{j2}$ of ultrasonic signal is also recorded. If speed of the ultrasonic signal is $u$, an estimation of the distance $d$ between the beacon transmitting node and the receiving dumb node is given by:

$$d \approx (t_{j2} - t_{j1}) \times u$$

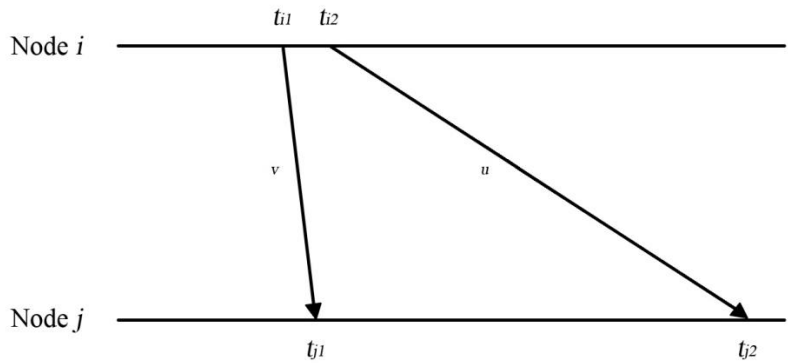

Figure 3. Signal propagation times in multi-signal TDOA distance estimation technique.

It should be noted that sender and receiver clocks need not be synchronized because the RF signal provides for an indirect synchronization mechanism between the transmitting and receiving nodes. Disadvantage of this scheme is that the nodes must be equipped with separate pieces of hardware for transmission, reception and processing two different kinds of signals. For the ultrasonic signals, the nodes must be equipped with microphones and speakers. This results in increased cost, size and energy consumption in the sensor nodes. Instead of RF and ultrasonic signals, any two signals with large difference in their speeds can be used for multi-signal TDoA. An example of localization system using multi-signal TDoA is the Cricket location support system [18][19], which is described later in this chapter.

Position estimation using multi-signal TdoA is quite accurate compared to other ranging techniques. Priyantha, Miu, Balakrishnan & Teller [20] have reported achieving accuracy to the order of a few centimeters. However, this is only possible under line of sight conditions. Under non line of sight conditions, both RF and ultrasonic signals suffer from propagation delays, and position estimation using multi-signal TDoA is error prone. In particular, ultrasonic waves may suffer from attenuation due to scattering and diffraction resulting from the obstacles and atmospheric effects such as temperature, pressure, humidity and turbulence.

## 3.4 Angle of Arrival

The direction of arrival of a signal at the dumb node can also be used to estimate its position. The direction of a received signal can be determined by measuring the angle it makes with some reference direction or orientation. Alternatively, the angle between the dumb node and the beacon node may be measured. For the localization of a dumb node using this technique, angles of arrival from a minimum of three beacon nodes are measured. Position information of three or more beacon nodes along with the three angles of arrival can be used to estimate the location of the dumb node.

The angle of arrival can be measured using directional antennas, a special configuration of antenna arrays or a combination of both. When using directional antennas, these can be mounted on the beacon nodes. To serve multiple dumb nodes, a directional antenna mounted on a beacon node rotates about its axis thereby transmitting beacon signals in all directions. A dumb node may use a similar directional antenna configuration to receive the beacon signals. Alternatively, dumb nodes can also use special configuration of antenna arrays to receive and measure the angle of arrival of a beacon signal. When an antenna array is used, antennas in the array are placed at known separation. The difference of time of arrival of the wave front at different antennas is used to estimate the direction from which the signal arrived.

Practical use of this technique is limited due to the complexities of deployment of special antennas. For example, mounting rotating directional antennas on tiny nodes is problematic and the rotating components are more prone to failure. Similarly, if an antenna array configuration is used, antennas in the array must be placed specific distance apart which is again a difficult proposition considering the tiny sizes of sensor nodes. Moreover, a greater accuracy of angle measurement is achieved only when separation distance between antennas in the array is small. However, with smaller separation distance, more sophisticated and precise hardware is needed for time difference measurements. Furthermore, shadowing, multipath fading and non line of sight conditions introduce a large amount of error in the estimated position which is more than same kind of errors in other similar techniques e.g. RSSI, ToA and TDoA. Due to these reasons, angle of arrival is considered less of a choice for localization in sensor networks.

After distances have been estimated by using one of the techniques discussed above, multilateration is employed to estimate the position of a dumb node. Obviously these techniques form the basis of range-based localization algorithms. Range-free algorithms do not use measurements to estimate distances and for localization. Instead, range-free algorithms analyze the connectivity information of neighbor nodes to deduce position information.

Apart from range-based and range-free methods, still another possible technique of localization is signal pattern matching. A database of unique signal signatures for all possible locations is created by using some property of radio signals. A dumb node localizes itself by comparing the pattern of received signals with the stored signal signatures. For example, a fixed number of beacon nodes may be deployed in the sensor field and RSSI values at each possible location may be calculated and stored in a database so as to serve as location signatures. This database can then later be used for localization after actual sensor nodes are deployed in the sensor field. It should be noted that other signal characteristics instead of RSSI values, such as multipath pattern of a signal arriving at a given location can also be used to create unique signatures for each location. Similarly, multiple signal characteristics may be combined to develop the signature database. This information can then be used to locate the position of a node. However, due to its very nature, this technique is not suitable for networks which are ad hoc and may have a dynamically changing topology, which is the case with sensor networks.

## 4. SINGLE-HOP LOCALIZATION SCHEMES

A number of localization schemes for single-hop networks have been developed which employ the principles and techniques discussed in the previous section. In a single-hop network, a dumb node has a direct i.e. single-hop communication link to the beacon nodes. Some of the localization schemes for single-hop networks include active badge, active office, cricket, connectivity based centroid algorithm and APIT. Most of these single-hop localization algorithms were developed for context-aware computing before research on wireless sensor networks gained focus.

In a wireless sensor network, a dumb node may not always have a direct communication link with a beacon node. In other words, a wireless sensor network is usually multihop in nature. Therefore, localization schemes for single-hop networks are not suitable for large multihop wireless sensor networks. However, they provide insight into the problem of localization and provide basic principles, techniques and foundation to develop algorithms for multihop sensor networks.

### 4.1 Active Badge

Active badge location system is one of the earliest location estimation systems proposed by Want, Falcao and Gibbons [32]. It is meant to find location information of people in a building. Each person to be localized is supposed to wear a badge, which is termed *active badge*, because it emits an infrared beacon signal carrying a unique identification number after every 15 seconds. The beacon signals transmitted by the active badges are collected by sensors which are placed at appropriate known locations throughout the building e.g. one such sensor may be installed in every room. The sensors, in turn, are connected to a central computer which processes the information received from these sensors and determines position of badges and hence persons wearing them.

The active badge localization system is range-free and uses centralized processing to determine locations of badges. The location information provided by the active badge scheme is symbolic i.e. it is a coarse-grained scheme which provides location information of a person, say, as in room 34.

Badges and sensors use pulse-width modulated (PWM) infrared (IR) waves for signaling. Unlike radio signals, IR signals do not penetrate walls and partitions in a building. Instead these are reflected and are not directional inside small rooms. Due to this property, location of a badge inside a certain room can be determined. Disadvantage of using infrared is that the badge has to be worn by the persons to be localized.

Variations of active badge scheme have been proposed for other applications. For example, Harter and Hopper have suggested the use of an *equipment tag* [12], similar to active badge for the location estimation of equipment and devices for context-aware applications.

**4.2 Active Office**

The active office localization technique is range-based which uses central processing and provides fine-grained localization. It was proposed by Ward, Jones and Hopper [33] to determine positions of devices for context-aware applications. The scheme uses multi-signal TDoA to estimate distances. Each device, which is to be localized, carries a wireless transceiver unit consisting of a microprocessor, 418 MHz radio transceiver, and an array of ultrasonic transducers. Each of these units has a unique 16-bit address. On the ceiling of the room in which positions of the devices are to be determined, a matrix of ultrasonic receivers is mounted. These receivers are also connected to a central computer through a serial link as shown in Figure 4. The central computer also controls an RF transmitter unit which broadcasts a radio message consisting of a preamble and 16-bit address every 200 ms. The central computer chooses the device whose 16-bit address is to be sent in the next message. At the time radio message is broadcast to the devices to be localized, a reset signal is also sent to the ultrasonic receivers mounted on the ceiling through the serial link.

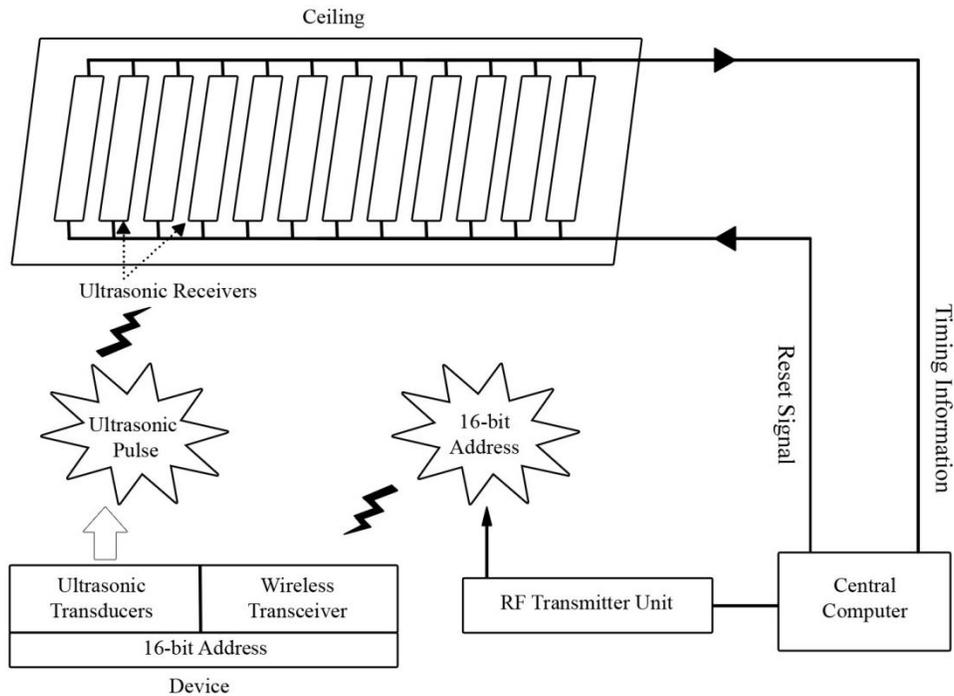

Figure 4. Device localization using active office technique.

All the devices receive the broadcast radio message. The device, whose address matches the address in the message, transmits a short ultrasonic pulse, which is received by the ultrasonic receivers mounted on the ceiling. The receivers measure the time of flight of ultrasonic pulse using the difference in time when it received the reset signal from the time it received the ultrasonic pulse. The timing information is then used to estimate distance of the device from the ultrasonic receivers. As positions of receivers are already known, distance information from these receivers can be used to solve a multilateration problem to localize the device.

A very good accuracy of at least 95% of averaged readings within 8 cm of true position has been reported. Authors claim that if enough transducers are mounted on the device, the accuracy is high enough to provide information relating to the orientation of the device as well.

**4.3 Cricket**

In the active badge and active office location estimation systems described previously, location is calculated by a central computer. This information is not communicated to the badge or device itself, and it is supposed to be used by the central computer for context-aware applications. In the Cricket location support system [19] proposed by Priyantha, Chakraborty and Balakrishnan, devices calculate their own positions instead of a central computer. The design goals of Cricket include user privacy, decentralized administration, network heterogeneity, low cost and fine-grained localization. Like active office system, Cricket uses multi-signal TDoA with a combination of radio and ultrasound signals for distance estimation and thereby position determination. The reported location granularity achieved by the Cricket system is $4 \times 4$ ft$^2$.

Rather than mounting it on the devices to be localized as in active office scheme, the Cricket system places RF transmitter and ultrasonic transducer in the ceiling of the room so as to serve as beacon. The device to be localized serves as listener and performs local processing to determine its own position. Multiple beacons are placed in the ceiling of the room of interest. Each beacon transmits an RF and ultrasonic signal concurrently. The device receives the RF signal first and then waits for the ultrasonic pulse to arrive. Time difference between the arrivals of RF and ultrasonic signals is used to estimate the distance from the beacon node. Distance and position information from three or more beacon nodes enables the device to determine its own position by solving a multilateration problem.

In the Cricket localization scheme, the nodes are not time synchronized. As a result, RF transmissions from different beacon nodes may collide. Furthermore, a listening device may correlate RF signal from one beacon node with ultrasonic pulse from another beacon node. This problem can be solved by using a carrier-sense style channel access protocol. However, it would increase cost and energy consumption. An alternative is to use fixed or deterministic transmission schedule for the beacon nodes. However, this is also prone to problems due to clock drifts. Cricket system, instead, uses randomization to alleviate the problem of beacon collisions. It chooses the beacon transmission times in a random manner with uniform distribution within a given interval. As a result, broadcasts from different beacons are independent so as to avoid repeated synchronization and at the same time prevent persistent collisions.

## 4.4 Bulusu's Algorithm

Unlike the single-hop localization schemes discussed above which were basically designed for context-aware applications, Bulusu's algorithm was designed for the localization of nodes in sensor networks. This single-hop, range-free localization algorithm due to Bulusu, Heidmann and Estrin [6] assumes an idealized radio model where all sensor nodes possess the same transmission power so that their transmission range is identical. It is further assumed that all sensor nodes transmit in an idealized spherical fashion. Each node transmits periodic beacon signals every $T$ seconds containing its position information. The neighbor beacon nodes are synchronized in time such that their beacon signals do not overlap in time, and in any time interval $T$, each beacon node transmits exactly one beacon signal. However, beacon nodes have an overlapping region of transmission.

Each dumb node $j$ keeps a count of the number of beacon signals received from a particular beacon node $i$ in some fixed time interval $t$. Knowing the time period $T$ after which a beacon signal is transmitted by the beacon node $i$, the dumb node $j$ can also compute the total number of beacon signals transmitted by the beacon node $i$ in time interval $t$. Using both these parameters, the dumb node $j$ can compute a connectivity metric for a particular beacon node. The connectivity metric is given by the percentage of beacon signals received by the dumb node $j$ which were transmitted by the beacon node $i$. If

$N_s(i, t)$ = Number of beacons transmitted by beacon node $B_i$ in time $t$. Subscript $s$ in $N_s(i, t)$ refers to *sender* node.

$N_r(i, t)$ = Number of beacons transmitted by beacon node $B_i$ and received by dumb node $D_j$ in time $t$. Subscript $r$ in $N_r(i, t)$ refers to *receiver* node.

$CM_{ij}$ = Connectivity metric for beacon node $B_i$ at dumb node $D_j$.

Then

$$CM_{ij} = \frac{N_r(i,t)}{N_s(i,t)} \times 100$$

Higher is the value of $CM_{ij}$, greater is the number of $B_i$ beacon signals received by the dumb node $D_j$, and smaller is the distance between them. The dumb node $D_j$ calculates this connectivity metric for all the beacon nodes in the set $S$ comprising of all beacons nodes from which it receives beacons. From this set $S$ of beacon nodes, it selects a subset $N$ of those neighbor beacon nodes for which the connectivity metric exceeds a certain threshold, say, 90 percent. The dumb node $D_j$ then localizes itself by determining the centroid of the selected beacon nodes. If the number of beacon nodes selected in subset $N$ is $k$ and their positions are given by $(x_{i1}, y_{i1}), (x_{i2}, y_{i2}), \ldots (x_{ik}, y_{ik})$, then estimated location of the dumb node $D_j$ is given by:

$$(x_j, y_j) = \left(\frac{x_{i1} + \ldots + x_{ik}}{k}, \frac{y_{i1} + \ldots + y_{ik}}{k}\right)$$

According to experimental results obtained by Bulusu, Heidmann and Estrin [6], the localization error falls within 30 percent of the separation distance between two adjacent beacon nodes. One major advantage of Bulusu's algorithm and all other proximity based and range-free algorithms is that no additional hardware is required to measure timing information so as to calculate distance between beacon node and dumb node. Without this additional hardware, size of the node remains tiny, and energy, which is a scarce resource in sensor networks, is also conserved.

Disadvantage of Bulusu's algorithm is that it requires a high density of beacon nodes in the sensor field so that each dumb node has at least three neighbor beacon nodes. Furthermore, the algorithm works only for single-hop sensor networks and is not suitable for sensor networks which are multihop in nature.

## 4.5 Approximate Point In Triangulation (APIT)

APIT is an area-based range-free localization scheme proposed by He, Huang, Blum, Stankovic and Abdelzaher [13]. Authors of this scheme proclaim that it performs best in a sensor network with an irregular radio pattern, random node placement and has a low communication overhead. Rather than context-aware applications, APIT algorithm is designed for sensor networks. APIT uses beacon nodes and RSSI information from neighbor nodes of a dumb node to solve the problem of localization. It employs distributed processing and each dumb node determines its position by locally processing the available information.

At first, three beacon nodes at a single hop from the dumb node are chosen and the dumb node determines whether it is inside or outside the triangle formed by the three beacon nodes. The decision whether the dumb node is inside or outside the triangle is made by using the RSSI information from neighbor nodes of the dumb node. Next, a different set of beacon nodes is selected and again the dumb node determines whether it is inside or outside the triangle formed by the new set of beacon nodes. After repeated iterations of this process, the dumb node is then able to estimate its position by narrowing down the estimate of the area of intersection of the triangles it was found to be inside and excluding the area of triangles it was found to be outside. The process of repeated selection of beacon nodes and triangle test is carried out until all the audible beacon nodes are exhausted or the required level of accuracy of location information is achieved.

The decision whether a node is inside or outside a triangle is based upon, what the authors of the scheme call, Point-In-Triangulation Test (PIT). PIT test is further categorized as perfect PIT test and Approximate PIT test (APIT). According to the perfect PIT test theory, if a node M is inside a triangle, when M is moved in any direction, it must be nearer to or farther from at least one vertex of the triangle. Similarly, if the node M is outside a triangle, when M is moved, there must exist a direction in which the new position of M is closer to or farther from all the three vertices of the triangle. The concept is illustrated in Figure 5.

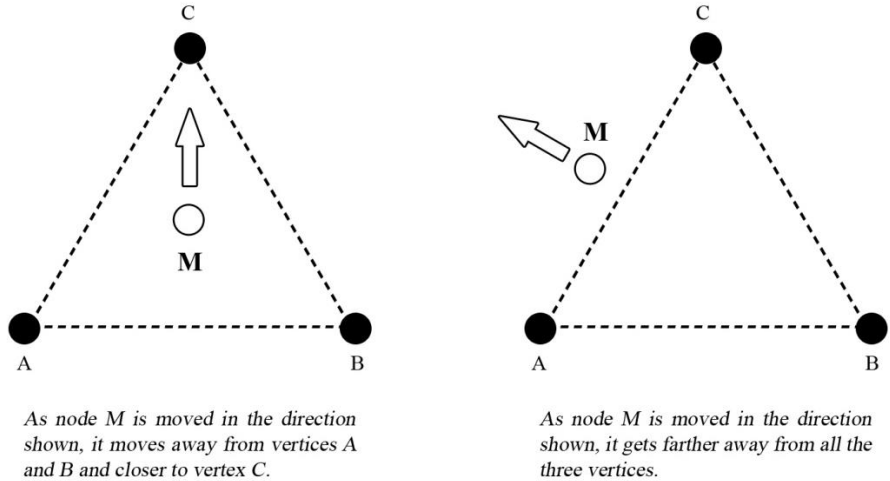

| As node M is moved in the direction shown, it moves away from vertices A and B and closer to vertex C. | As node M is moved in the direction shown, it gets farther away from all the three vertices. |

Figure 5. Perfect PIT Test.

In a sensor network comprising of static sensor nodes, it is not possible to move nodes. Therefore, an approximation of the Perfect PIT Test called Approximate PIT Test (APIT) is used instead. In the approximate PIT Test, a dumb node uses information from other neighbor dumb nodes to determine whether it is inside or outside a triangle. If no neighbor of a node M is farther from or closer to all the three vertices of a triangle simultaneously when compared to the node M, then node M assumes that it is inside the triangle else it assumes that it is located outside the triangle as is illustrated in Figure 6.

A dumb node can decide whether a neighbor node is farther from or closer to all the three vertices of a triangle than itself by comparing the RSSI information from the neighbor node and comparing it with that of its own. A higher RSSI value of a neighbor dumb node than that of the host dumb node received from the same beacon node means that the neighbor node is closer to the beacon node than the host node itself.

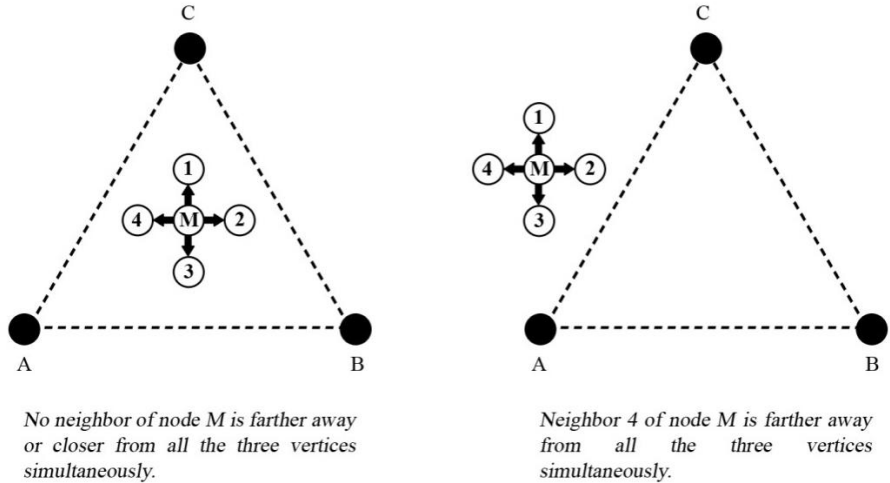

Figure 6. Approximate PIT Test.

The assumption that the distance of a dumb node from a beacon node is proportional to the RSSI value is a source of error in APIT because RSSI values are also affected by such factors as non line of sight conditions, reflection and scattering. As a result, a dumb node which is otherwise closer to a beacon node may have a smaller RSSI value due to obstacles. Similarly, RSSI values may also vary due to burst disturbance effects. However, error due to it can be minimized by taking a running average of RSSI values over time.

There are other possible sources of error in APIT as well. For example, consider a dumb node M, which lies inside a triangle closer to one of its edges. If a neighbor N of the dumb node M lies outside the triangle near the same edge at a greater distance, than it is farther away from all the three vertices (beacon nodes) of the triangle and hence node M can also assume that it, too, is outside of the triangle.

Severity of errors discussed above can be reduced by making more PIT tests, which is possible only if there is high density of beacon and sensor nodes in the sensor network. Therefore, APIT performs better only in dense sensor networks.

## 5. MULTIHOP LOCALIZATION SCHEMES

In practical applications of wireless sensor networks, majority of dumb nodes are located at a multihop distance from the beacon nodes. Fraction of dumb nodes located at single hop or multihop distance from a particular number of beacon nodes depend upon a number of factors such as density of beacon nodes, size of sensor field, nature of the application for which sensors are deployed and topology of the sensor network. The single-hop localization schemes discussed previously are not applicable under these circumstances.

Using the same principles and techniques of distance estimation and localization as discussed in the previous section, a number of localization algorithms for multihop wireless sensor networks have been developed. Some of these algorithms are only academic in nature whereas some are more suited to practical applications. In this section we describe a few of the representative localization algorithms for wireless sensor networks.

## 5.1 Ad Hoc Positioning System

One of the earliest localization algorithms for multihop sensor networks that estimates node positions from mere connectivity information, and, hence, is range-free, is Ad Hoc Positioning System (APS) [16], which was proposed by Niculescu and Nath.

The algorithm uses distributed processing and each dumb node determines its position using multilateration. To perform multilateration, range estimates to at least three beacon nodes are required. However, in a multihop sensor network, it is not possible for each dumb node to have a direct single-hop communication link with three beacon nodes. To resolve this problem, APS proposes to use connectivity information to estimate range of a beacon node which is at a multihop distance from the dumb node. It suggests three basic techniques to perform range estimation of multihop distant beacon nodes via intermediate nodes:

- DV-Hop Propagation Method
- DV-Distance Propagation Method
- Euclidean Propagation Method

All three techniques use flooding to propagate information in the network similar in nature to the operation of distance vector (DV) routing protocol. Starting with each anchor, a node propagates information to its immediate first-hop neighbors only. The propagated information depends upon which of the three variation of the APS algorithm is being used. This approach is well suited to nodes having limited bandwidth and power. It is to be noted that level and complexity of signaling depends on the total number of beacon nodes and average degree of each node i.e. number of single-hop neighbors of a node.

Each of the three techniques listed above vary in the amount of signaling, power consumption and the degree of position accuracy achieved and hence each technique may be suitable for a certain class of problems.

*5.1.1 DV-Hop Propagation Method*

This is the most basic form of APS algorithm. Each node in the sensor field maintains a table having entries of the form {$X_i$, $Y_i$, $h_i$} where ($X_i$, $Y_i$) is the location of a beacon node $i$, and $h_i$ is the number of hops between node maintaining the table and the beacon node $i$. The number of hops to a beacon node can be counted by incrementing a *count* field in a message at each hop as it is transmitted from a beacon node to its immediate neighbors and so on. This table is maintained by the beacon node $i$ itself as well. When the beacon node $i$ has obtained positions and hop counts for all other beacon nodes $j$, it can then calculate average size of a hop as under:

$$c_i = \frac{\sum \sqrt{(X_i - X_j)^2 + (Y_i - Y_j)^2}}{\sum h_i}$$

for all beacon nodes $j$ where $i \neq j$.

This average size of one hop, $c_i$, computed by the beacon node $i$ is treated as a correction factor and is propagated throughout the network using controlled flooding as discussed earlier. Knowing the locations of beacon nodes and the correction factor $c_i$, a dumb node can then use multilateration to estimate its own position. The steps involved in the APS DV-Hop algorithm can be summarized below:

- The algorithm takes its start from the beacon nodes which propagate their position information to their immediate neighbors.

- All other nodes work along the same technique as distance vector algorithm and receive and propagate the position information of beacon nodes. Ultimately, all the nodes have position information of all the beacon nodes and also the number of hops to these beacon nodes.

- When a beacon node has received the position of other beacon nodes and the number of hops to them, it can then compute the average size of one hop.

- The average hop size is propagated as correction factor to all other nodes in the network using flooding in a controlled manner.

- All dumb nodes already have number of hops to the beacon nodes. Now, when a dumb node receives the correction factor, it can convert its distance to the beacon node from hops to units of length.

- Finally, multilateration is used to compute the node position.

It should be noted that the correction factor calculated by each beacon node will be different. Therefore, each dumb node will receive different correction factors from different beacon nodes. Niclescu and Bath suggest that a dumb node should use the first correction factor it receives to estimate its position and drop all the subsequent correction factors. In general, such a policy will ensure that a dumb node should use the correction factor received from its closest beacon node. In addition, if the network is large, a TTL field can be set in the packets used for the distribution of correction factors so that amount of signaling and congestion in the network is reduced. This will again ensure that the correction factor calculated by a beacon node is used by the dumb nodes in its immediate neighborhood. Limiting the propagation of correction factor by TTL field also supplements the policy of usage of the first correction factor received.

Distance calculation and position estimation by using the correction factor in DV-Hop propagation method is best explained with the help of an example. Consider segment of a sensor network with three beacon nodes, $B_1$, $B_2$ and $B_3$ and six dumb nodes as shown in the Figure 7. Each of the beacon nodes knows the Euclidean distance to other beacon nodes as given in the Figure. Now, the correction factors computed by $B_1$, $B_2$ and $B_3$ are:

$$c_{B1} = \frac{60 + 150}{2 + 5} = 30 \text{ m}$$

$$c_{B2} = \frac{60 + 100}{2 + 4} = 26.667 \text{ m}$$

$$c_{B3} = \frac{150 + 100}{5 + 4} = 27.778 \text{ m}$$

Let us now consider dumb node $D_1$ and find out the distances it estimates to the three beacon nodes. $D_1$ is at a distance of 3 hop counts from the beacon node $B_1$, 2 hop counts from $B_2$ and 3 hop counts from $B_3$. With 2 hop counts, $B_2$ is the nearest beacon and most likely, it will receive the first correction factor i.e. 26.667 m, from this beacon node. As a result, the dumb node $D_1$ can now estimate distances from the three beacon nodes as under:

Distance of $D_1$ from $B_1$ = 26.667 x 3 = 80 m
Distance of $D_1$ from $B_2$ = 26.667 x 2 = 53.334 m
Distance of $D_1$ from $B_3$ = 26.667 x 3 = 80 m

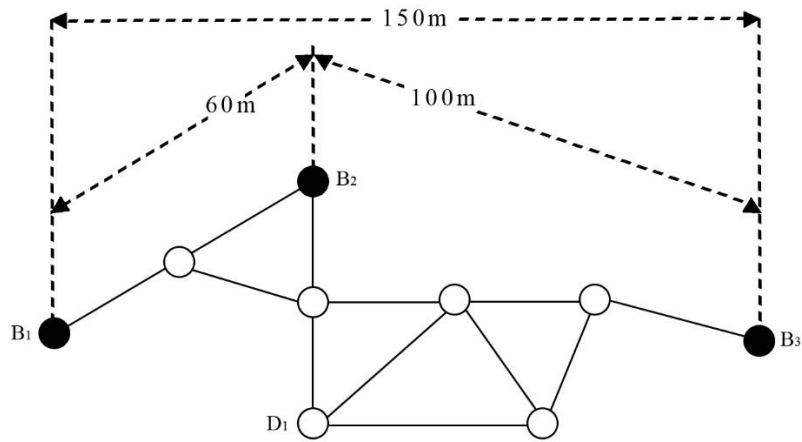

Figure 7. Node localization using DV-Hop propagation method.

$D_1$ already knows the position information of the three beacon nodes. Using the additional information of distances to these beacon nodes, it can now employ trilateration to estimate its own position.

Main advantages of the DV-Hop propagation method are simplicity, smaller amount of processing required, lower cost as no additional equipment on sensor nodes is required, low energy cost and the fact that measurement errors are avoided as it does not use any measurement technique for range estimation. Its major drawback is that it is most suitable only for isotropic networks where hop length is uniform across all segments of the network. The algorithm may not produce desired results for anisotropic networks, where node density and hence hop length in certain segments of the network does not correspond well to the correction factor calculated by the beacon node.

## 5.1.2 DV-Distance Propagation Method

In this variation of the APS algorithm, range between neighbor nodes is estimated using RSSI measurements and this distance information instead of average size of hop is propagated in the network. In other words, the distance vector algorithm now uses cumulative distance instead of hop counts. As a result, a dumb node now knows its distance from the beacon nodes by receiving this distance information through DV exchange instead of computing it by multiplication of hop counts with average size of a hop as in DV-Hop. Knowing the distance information and positions of beacon nodes, a dumb node uses multilateration to estimate its own position.

As the distance estimates are better in this technique, DV-Distance method provides better position information than DV-Hop i.e. position information calculated by DV-Distance is less coarse. Furthermore, DV-Distance method is suited to anisotropic networks as well, because the variation in the size of hops between various nodes does not affect the actual measured distances between the nodes as is the case with DV-Hop propagation method. However, this method has its own drawbacks, e.g. measurement errors can produce incorrect results.

*5.1.3 Euclidean Propagation Method*

This method provides better and fine-grained position information compared to DV-Hop and DV-distance methods. It works by propagating the true Euclidean distance to the beacon nodes. Combining this distance information with the positioning information of beacon nodes, a dumb node is able to estimate its own fine-grained position using multilateration.

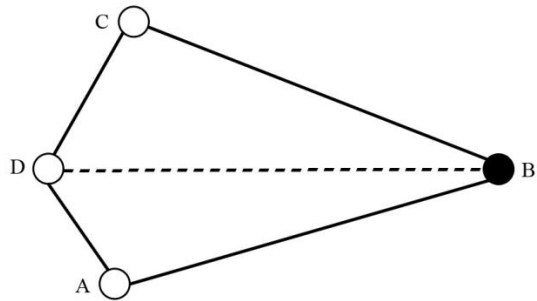

Figure 8. Node localization using Euclidean propagation method.

For a dumb node D to be able to calculate its Euclidean distance from a beacon node B, it must have at least two neighbors A and C which have distance estimates for the beacon node B. In Figure 8, the neighbors of D i.e. A and C have distance estimates AB and CB. The dumb node D also has estimates for AD, CD and AC. With this available information, the dumb node D can calculate the Euclidean distance BD from the beacon node B. In a similar manner, when node D has calculated Euclidean distances to three or more beacon nodes, it can use multilateration to estimate its own position.

This method is also suitable for anisotropic networks. However, it requires a higher ratio of beacon nodes compared to DV-Hop and DV-Distance methods which work well with smaller number of beacon nodes.

## 5.2 Convex Position Estimation

Convex Position Estimation (CPE) algorithm was proposed by Doherty, Pister and Ghaoui [10]. It treats the problem of sensor node localization as an optimization problem and applies linear programming to arrive at the solution for the localization problem.

A convex set is a set of points such that any two points in the set can be connected with a path comprising of points contained within the set. The set of all the nodes in a sensor network forms a convex set as a path between any two nodes comprises of only nodes which are within the set of the sensor nodes. There is a path P between nodes M and N if either of M and N communicates with the other or both M and N communicate with each other. The path comprises of nodes M and N and all other intermediate nodes which participate in communication between them.

A linear program (LP) is a mathematical optimization technique to find out the best possible solution to a problem satisfying a set of linear constraint relationships. For example, an LP problem can be formulated to find the minimum labor needed to complete a given task under certain constraints. Formally, an LP problem can, in general, be represented as:

Minimize $\quad c^T x$
Subject to $\quad Ax \leq b$

where $x$ is the vector of variables to be determined, and $c$ and $b$ are vectors of known values. $A$ is a matrix of known values. The expression to be optimized i.e. $c^T x$ is called the object function. The relational expression $Ax \leq b$ specifies the constraints which should be satisfied by the optimized solution.

Convex position estimation algorithm provides an optimization framework and formulates the problem of position determination of dumb nodes as linear program. For a two-dimensional sensor field, each node's position can be described in rectangular coordinates as ($x$, $y$). For the LP problem, a single vector $x$ with positions of all the nodes is formed:

$$x = [x_1y_1 \ldots x_my_m, x_{m+1}y_{m+1} \ldots x_ny_n]$$

The first *m* entries are known positions of beacon nodes and the remaining entries *n-m* are determined by the algorithm.

As is clear, locations of a subset of nodes i.e. beacon nodes are known. Furthermore, connectivity information of beacon and dumb nodes is also known which is modeled as a set of geometric constraints on the node positions. For example, if a node M can transmit in a radius of 10 meters, and another node N can receive communication from node M, then it means that it lies within 10 meter radius of node M. Similarly, another node P which does not receive communication from node M, lies outside of this 10 meter radius around node M. This connectivity information serves as constraint for the linear program problem to be solved for the determination of node positions. Such connectivity constraints limit the set of estimated positions of dumb nodes. In general, the localization problem is then formulated as under:

Given:      Positions of beacon nodes
Find:       Positions of dumb nodes
Subject to: Connectivity constraints of the beacon and dumb nodes

Each node in the sensor field maintains a table of connectivity information by exchanging messages with the neighbor nodes. The table maintains a list of neighbor nodes from which the node receives messages. A receiver node can receive messages from only that transmitter node whose transmission radius encompasses the receiver node. This provides a natural bound on the distance between the transmitter and receiver nodes.

Each node sends its table of connectivity information i.e. list of neighbor nodes from which it receives messages to a central processing device. The central processor also has the list of all beacon nodes along with their position information. The central processor then solves the linear program problem to determine positions of the dumb nodes. Using the connectivity constraints and knowledge of positions of the beacon nodes, the algorithm narrows down the area and generates a bounding box where a dumb node may be located. The bounding box is treated as a set of all possible locations where the dumb node may be located. The algorithm, then, selects the center of the bounding box as the best possible estimate of the node position.

Higher is the number of beacon nodes, smaller is the size of the bounding box thereby giving a better estimate of the node position. If the number of beacon nodes used in defining the bounding box is small, the *settled nodes* – the dumb nodes whose positions have already been determined, can also be used along with the beacon nodes to find positions of remaining dumb nodes. However, location estimates of the settled nodes can have errors, which may propagate in subsequent computations and hence result in inaccurate position estimates of the dumb nodes. To overcome this problem, authors of the algorithm suggest the use of eight settled nodes to generate the same certainty of position estimation as given by a single beacon node.

Additional constraints, such as angle of arrival, can also be used along with connectivity information to generate a smaller bounding box, and as a result, a better estimate of the node position can be obtained. As the CPE algorithm does not require range estimation between nodes, it is a range-free algorithm. As a result, it does not require any additional hardware on the sensor nodes. However, it is a centralized algorithm, and all the nodes must send connectivity information to the central processor. The central processor, after processing and determination of node positions, sends position information back to each node. This entails communication overhead resulting in high energy cost. Security of node position information may also be compromised during processing at the central node or during transmission from central processor to the dumb node. As in all centralized algorithms, the central processor can also be a single point of failure.

Due to these problems, CPE algorithm may not be a viable option of localization in current practical applications of sensor networks. However, it provides an additional insight and a method to approach the solution to the localization problem and can serve as a platform to develop and optimize better localization algorithms.

## 5.3 Iterative and Collaborative Multilateration

Once a dumb node has estimated its position with the help of beacon nodes, it can be used as a beacon node by other dumb nodes to estimate their positions. This basic scheme of position estimation is the basis of iterative location estimation [22][24]. The algorithm is essentially a two-step process. As proposed by Savarese, Rabaey and Beutel [22] and Savarese, Rabaey and Langendoen [23], all the dumb nodes compute a rough estimate of their positions in the first phase using an algorithm called Hop-Terrain. This is referred to as start-up phase. In the second phase, called the refinement phase, the computed position estimates are refined using an iterative process in which the immediate neighbor nodes of a dumb node take part in the process.

A similar algorithm has been proposed by Savvides, Han & Strivastava [24]. The authors call this algorithm as Ad-Hoc Localization System (AHLoS). The algorithm builds upon the idea of atomic multilateration. The case where a dumb node is at a distance of one hop from at least three non-collinear beacon nodes, the dumb node can perform multilateration to determine its position. This is termed as atomic multilateration. The iterative multilateration uses the atomic multilateration as its main primitive. The algorithm supports central, distributed or cluster-based processing. If the algorithm is deployed using distributed processing, first the dumb nodes with a distance of one hop from at least three beacon nodes estimate their position using atomic multilateration. These dumb nodes, after their positions have been estimated, are immediately available as beacon nodes and the other dumb nodes can then utilize their position information for their own position estimation using iterative multilateration.

If the algorithm uses centralized processing, the central node has the global knowledge of the sensor network. The network is represented by a graph G on which the algorithm operates. The graph has weighted edges to denote the separation between two adjacent nodes. At first, the positions of those dumb nodes which are at a distance of single hop from at least three beacon nodes are determined using atomic multilateration. With the central processing, the algorithm first determines positions of those dumb nodes that have connectivity with maximum number of beacon nodes. This results in better accuracy and faster convergence. As soon as a dumb node has settled i.e. it has estimated its position, it becomes a beacon node so that remaining dumb nodes can use it for multilateration. This iterative procedure repeats until the positions of all the dumb nodes have been estimated.

A drawback of the iterative multilateration is the propagation and accumulation of error which results from the use of settled nodes as beacon nodes. As the position estimates of these nodes are not as precise as beacon nodes, the multilateration involving these nodes may accumulate these errors.

It is also possible that the algorithm does not converge due to accumulation of error and the accumulated error increases with time if the algorithm keeps running. As a countermeasure, it has been suggested to add confidence weights to position estimates of settled nodes being used as beacon nodes and then solve a modified optimization problem [23]. With this approach, the algorithm converges for almost all cases of location estimation and also produces better position estimates. Furthermore, the convergence problem can also be avoided by using better and high precision ranging techniques.

For iterative multilateration to be possible, at least three beacon or settled nodes must be in the neighborhood of the dumb node for which position is to be estimated. However, this is not always the case. Not all the nodes in the sensor field have three neighbor nodes with known position estimates. Under such situation, it is not possible to find the position of a dumb node by mere atomic or iterative multilateration. Another technique called collaborative multilateration [23][24] may be used for position estimation of such nodes. Collaborative multilateration is a technique in which a dumb node uses multi-hop information from another neighbor dumb node and solves a joint set of location estimation functions to determine positions of both dumb nodes simultaneously. The solution obtained may or may not be unique. A unique solution to the problem of location estimation using collaborative multilateration is reached only when a participating node has at least three participating neighbors. A *participating node* is a sensor node which is either a beacon node or a dumb node with at least three participating neighbors.

Collaborative multilateration may be used as an enhancement to iterative multilateration. As a result of collaborative multilateration, the number of dumb nodes reduces and the number of settled nodes increases. These settled nodes can further be used in another round of iterative multilateration to determine locations of more nodes and to improve the accuracy of location estimates.

### 5.4 Multidimensional Scaling

Multidimensional analysis is a data analysis technique used primarily in psychometrics and psychophysics. Shang, Ruml, Zhang and Fromherz applied this technique to propose a range-free localization algorithm MDS-MAP [27], which can estimate node positions using mere connectivity information. However, the algorithm can also make use of additional information e.g. position information of beacon nodes or estimated distances between neighbor nodes. With this additional information, the algorithm is able to provide better and accurate results. The algorithm works even if no beacon nodes are available in which case it generates relative coordinates for the dumb nodes.

The underlying theory and concept of multidimensional scaling (MDS) is best explained by an example described in [27]. Consider a small cloud of colored beads suspended in the air. The distances between each pair of beads are measured. Now, if the beads fall down on the ground, the earlier configuration of the beads in the air can be reconstructed by using the distance measurements recorded earlier. Position of each bead in the reconstructed arrangement is chosen such that the distances in the new arrangement match the distances in the original layout. It is this problem of reconstructing the arrangement that is solved by MDS.

It should be noted that the reconstructed arrangement will be an arbitrarily rotated and flipped version of the original layout because the new layout is constructed without using the absolute information.

MDS-MAP executes in three steps:
- Step – 1: Distance between each possible pair of nodes is estimated using an all-pairs shortest path algorithm e.g. Dijkstra's or Floyd's algorithm.

- Step – 2: Multidimensional scaling is used to estimate node positions such that the estimated positions satisfy the distances estimated in step 1. This step generates a relative map of the sensor nodes such that the nodes have same neighbor relationship as the underlying network. Relative map is generated in the absence of any beacon nodes and is well-suited for sensor networks in which powerful sensors and expensive infrastructure using beacon nodes cannot be installed. Relative positioning information is sufficient for some applications such as direction-based routing algorithms.

- Step – 3: Resulting coordinates are normalized by taking into account the known positions of beacon nodes. This step generates absolute position information of dumb nodes.

MDS-MAP is a centralized algorithm and is not much suited to practical applications of wireless sensor networks. An improved version of the MDS-based localization algorithm, called MDS-MAP(P), which is short for MDS-MAP using *p*atches of relative maps, has also been proposed [26][28]. Main theme of the improved MDS-MAP(P) is to build local maps of the neighbor nodes at each individual node and then to merge all the local maps to form a global map. In this way, computation of shortest path distances between far away nodes is avoided. Furthermore, the local maps of nearby nodes are usually more accurate.

The original MDS-MAP works well with sensor networks in which nodes are uniformly distributed and the shortest path between two nodes correlates with the true Euclidean distance between them. However, in the case of anisotropic and irregular networks, it suffers from performance degradation and the estimated node positions are prone to error. On the other hand, MDS-MAP(P) works well with both uniform and irregular networks. As it uses local maps, a good local map of far away nodes can be constructed and merged with the global map thereby providing better results for anisotropic networks. Another advantage of MDS-MAP(P) is that it offers partially distributed processing approach as local maps are constructed at individual nodes. However, a central processor is still required where the global map is constructed and the node positions are estimated.

**5.5 Miscellaneous Algorithms**

In addition to the algorithms described above, a number of localization algorithms have been proposed such as Recursive Position Estimation [3] and Directed Position Estimation [8]. Recursive Position Estimation (RPE), as the name suggests, utilizes recursive steps to estimate and refine the position estimation. Although its communication cost is low, it needs more beacon nodes. Directed Position Estimation (DPE) algorithm is like RPE but reduces the required number of beacon nodes by using directed recursion.

Examples of some recent localization algorithms are the ones proposed by Chang, Hung, Lin, & Li [7], Wang, Wu & Shu [30] and Rencheng, Lisha, Teng, & Liding [21].

Wang, Wu, & Shu have proposed a decentralized positioning system which utilizes a particle filter for more accurate measurement of Received Signal Strength (RSS) and then uses Least Mean Squares (LMS) algorithm to find the position of the node [30]. It is to be noted that accuracy of the location estimation depends upon how accurately the RSS is measured and the factors such as reflection, scattering and other similar phenomenon affect the recorded measurement. Therefore, the algorithm suggests the use of a particle filter to improve RSS measurement with subsequent application of Least Mean Square (LMS) algorithm to estimate the path loss factor of the distance-dependent path loss model. The algorithm is a range-based decentralized algorithm and hence requires less transmission energy when compared with a centralized approach.

Another low-cost strategy for localizing wireless sensor network nodes is bounding-box algorithm [11]. Whereas DV-Hop and range-based algorithms require static anchor, bounding-box algorithm does not require any fixed anchor and has low hardware requirements. This algorithm uses a mobile beacon node that is always aware of its position and broadcasts its current location while moving in the sensor field. The node receiving this beacon signal can then know the bounding-box region where it is inside and hence can find its own position. A problem with the bounding-box algorithm is that it is unable to locate relative positions of neighbor nodes. Due to this problem, bounding-box algorithm is not suitable for certain applications such as location aware routing. Chang, Hung, Lin, & Li have suggested a mechanism, which they call as Distinguishing Relative Locations or simply DRL to remedy this problem [7]. DRL uses mobile beacon nodes that transmit beacons to define the bounding-box region and tone signals to help distinguish relative locations of any two neighbor nodes. Unlike beacon signals, tone signals do not carry any position information and hence consume lesser amount of energy.

Another recent algorithm [21] combines two approaches i.e. iterative and collaborative multilateration using Maximum Likelihood Estimation [25] and MAX-MIN [29] localization algorithms. For small errors in distance measurement, the Maximum Likelihood Estimation (MLE) algorithm provides location estimation with reduced error. However, a drawback of the MLE algorithm is that it requires a lot of floating-point computation.

MAX-MIN algorithm is based upon bounding-box algorithm. After calculating the distance between the blind node and the beacon nodes, the MAX-MIN builds bounding boxes. The beacon nodes are taken as reference and circles are drawn with a reference node taken as the center. Radius of the circle is equal to the distance between the reference node and the blind node. Further, squares are circumscribed against each circle. Center of region of intersection of these squares is designated as the estimated location of the dumb node. We need at least three beacon nodes to locate a dumb node. MAX-MIN algorithm does not need complex computations and only simple additions and subtractions are required to estimate the position of a node. However, unlike MLE, the position accuracy is good enough for large errors in distance. If the distance error is below 7%, MLE estimates the position of blind node more accurately than the MAX-MIN algorithm. However, as the distance error increases beyond 7%, the MAX-MIN algorithm provides far more superior results than MLE algorithm.

The hybrid algorithm proposed by Rencheng, Lisha, Teng, & Liding combines the two algorithms to produce an optimum result. The hybrid algorithm at first estimates the location of the dumb node by using both MLE and MAX-MIN. Let ($x_1$, $y_1$) be the coordinates of a dumb node estimated using MLE and let ($x_2$, $y_2$) be the coordinates calculated using MAX-MIN. Then the hybrid algorithm is based upon choosing parameters $\alpha$ and $\beta$ so as to calculate the revised location with reduced error using the following formula:

$$\begin{bmatrix} x \\ y \end{bmatrix} = \alpha \begin{bmatrix} x_1 \\ y_1 \end{bmatrix} + \beta \begin{bmatrix} x_2 \\ y_2 \end{bmatrix}$$

Above formula combines results of both MLE and MAX-MIN to produce a location estimation which is better than results produced by either of the algorithms used alone.

## 6. FUTURE RESEARCH DIRECTIONS

Wireless sensor network is a new and emerging area of research. Practical applications of wireless sensor networks are still being conceived and developed. Along the way, new techniques and algorithms are being proposed and developed for various layers in the sensor networks. So is the case with node localization. As discussed earlier, majority applications of wireless sensor networks need to know the position information from where data was collected. This is possible only if the sensor node collecting and transmitting the data knows its position by using a localization algorithm.

Majority of proposed localization algorithms are generic in nature and do not take any particular application of sensor network into consideration. However, it is possible that a single localization algorithm may not be suited to the entire spectrum of wireless sensor network applications. For example, if an algorithm is suitable for the location awareness of sensor nodes in a body sensor network to monitor the physiological activity of a living being, it is much likely that it may not be suitable for the sensor network which is being used for the surveillance of a certain area such as a battlefield. Therefore, work needs to be done to determine the suitability of proposed localization algorithms for various applications, and if no current algorithm is suited to a particular problem, new algorithms might need to be developed and tested for it.

All the localization algorithms depend upon some kind of measurement, such as RSSI or timing, which is made by the underlying sensor hardware. However, these measurements are prone to errors due to practical limitations of the hardware and result in poor localization accuracy. This problem can be alleviated by careful calibration of sensor node or by making the algorithm robust against measurement errors by employing techniques to detect and either reject or correct these errors. Therefore, work is needed to determine the frequency of measurement errors, their impact on localization and how sensor nodes are to be calibrated so as to alleviate this problem. Similarly, new techniques need to be developed so that a localization algorithm is able to detect an improbable measurement. This can be accomplished, for example, by using consistency checks, such as, symmetry and geometric constraints. In a similar manner, statistical models may be used to filter out measurement errors.

Most of the current work in the area of node localization focuses on static sensor nodes as is the case with wireless sensor network applications. However, future applications will use mobile nodes as well. For example, a mobile node will be able to move to the area that needs sensing coverage. The localization algorithm should be able to detect this movement and determine the new position of the node. Similarly, a node move more frequently, for example, to collect data or transmit information to other nodes in the sensor field. The localization algorithm, in this case, should be able to determine the location of the moving node in real time. Major work needs to be done to resolve issues in the localization of such mobile nodes.

Some current and future applications involve a dynamically changing topology of wireless sensor networks with sensor nodes being added or removed from the sensor field. The localization algorithm should be scalable to either a very small or a very large number of sensor nodes and should provide the desired level of accuracy in both the extreme cases. Not many current localization algorithms take the scalability factor into consideration. So, additional work is needed so that the localization algorithms are scalable and work well with hundreds and thousands of sensor nodes.

Majority of node localization algorithms uses a set of beacon nodes whose position information is known either through GPS or similar device or by positioning them at locations with known coordinates. The beacon nodes know their positions throughout the period of their deployment. Work needs to be done to determine the minimum and optimum number of beacon nodes which will result in the desired accuracy of a localization algorithm. Similarly, some proposed localization schemes, such as localization using particle dynamics [36], use a large number of beacon nodes. By employing the principles used by these algorithms, new and efficient techniques which use fewer number of beacon nodes can be built.

Security of sensor nodes is an important issue in wireless sensor networks. Various types of attacks can be mounted against a sensor node at various layers. For example, a sensor node may be forced to provide wrong information by feeding it wrong input through a malicious sensor node planted by an adversary. Similarly, a dumb node may estimate wrong position for itself if an attacker is able to take control of a beacon node or is able to plant malicious beacon nodes in the sensor field. The compromised or planted malicious beacon node sends beacons signals with wrong position information to the dumb node. If the dumb node uses multilateration with this wrong position information as the input, the calculated position for the dumb node will be in error and, as a result, the data collected by this node will be ascribed to a wrong location by the sensor field. The problem can be alleviated if the nodes in the sensor field have a reliable mechanism to trust and authenticate each other. This and similar security issues need to be addressed in the localization algorithms.

The real world applications of sensor networks are deployed in a three dimensional space. However, the current work on localization algorithms focuses on two-dimensional planar space. Therefore, the current two dimensional localization algorithms need to be extended to three-dimensional space. Further experiments and simulations need to be carried out to test and verify their suitability for deployment in an actual three dimensional space.

Error accumulation and propagation is a severe problem in some range-free and range-based localization algorithms, such as, iterative and collaborative multilateration. Localization algorithm may not converge due to error propagation or it may result in unacceptable errors in location estimation. New techniques need to be developed to limit the accumulation and propagation of errors in localization so that the accuracy of localization can be increased.

## 7. CONCLUSION

In this chapter, we have defined the problem of localization of nodes in wireless sensor networks, described the desired characteristics of a localization algorithm and how these algorithms can be classified. We have also discussed a few of the representative localization algorithms. None of the current localization algorithms is suitable for the entire class of applications of wireless sensor networks. For example, a localization algorithm which is suitable for static sensor nodes may not work well with mobile sensor nodes. It should be noted that these algorithms describe only the basic principles and techniques which may be used for localization of sensor nodes. A complete localization application and framework for a practical wireless sensor network can be built using a combination of these techniques.

Each of these localization algorithms provides different level of accuracy of estimated positions and is directly related to the number of beacon nodes deployed in the sensor field. Better level of position accuracy can be achieved if a higher number of beacon nodes are deployed. Range-based localization schemes provide fine-grained position estimates. However, these techniques usually require extra piece of hardware to be added to sensor node for measurement of various parameters thereby adding to size, weight and energy costs. If size of integrated circuits and hardware keeps getting smaller following Moore's Law, range-based schemes might be able to use extra hardware without much additional cost and therefore may hold promise in future sensor networks. Range-free schemes, on the other hand, approximate the position of a dumb node by using mere connectivity and proximity information. Therefore, these schemes do not need extra hardware thereby saving cost in terms of weight, size and energy. However, location estimates provided by range-free schemes are coarse-grained.

Similarly, localization algorithms, such as MDS and MDS-MAP, may estimate only relative positions of nodes and other algorithms, such as APIT and APS, may localize using absolute coordinates. Localization algorithms which provide only relative positions do not need beacon nodes. However, if beacon nodes are also added to the network, these algorithms will determine absolute positions as well.

Algorithms with a centralized approach may have a large communication overhead and are not suited to the ad hoc nature of wireless sensor network. However, the accuracy of estimated positions is better because the central node has global knowledge of the sensor network. On the other hand, localization algorithms using distributed processing are easily scalable to small or large number of nodes in the sensor network.

Localization algorithms employing iterative and collaborative multilateration techniques hold promise as a large number of dumb nodes can be localized using fewer beacon nodes. If errors are detected and are prevented from being propagated, a good level of accuracy can also be achieved.

## 8. ACKNOWLEDGEMENTS

We would like to thank Mr. Mansoor Sheraz, software developer at Information Technology Center, COMSATS Institute of Information Technology, Lahore, for help in drawing the figures for this chapter.

## 10. ADDITIONAL READING